# On the Power-law Scaling of Turbulence Cospectra Part 1: Stably Stratified Atmospheric Boundary Layer

Yu Cheng • Qi Li • Andrey Grachev • Stefania Argentini • Harindra J.S. Fernando • Pierre Gentine

**Abstract** Turbulent fluxes in the atmospheric surface layer are key input for the prediction of weather, hydrology, and carbon dioxide concentration. In numerical modelling of turbulent fluxes, a -7/3 power-law scaling in turbulence cospectra is usually assumed at high wavenumbers. In eddy-covariance (EC) measurements of turbulent fluxes, an assumed shape of turbulence cospectra is typically required for high-frequency spectral corrections, typically assuming a -7/3 power law. The derivation of -7/3 power-law scaling is based primarily on dimensional analysis, and other cospectral scaling has also been observed. Here we examine the shape of turbulence cospectra at high wavenumbers from extensive field measurements of wind velocity, temperature, water vapour and $CO_2$ concentrations in various stably stratified atmospheric conditions. We propose a turbulence cospectral shape with -2 power law rather than -7/3 law for high wavenumber equilibrium

―――――――――――――――――――――――

Yu Cheng • Pierre Gentine
Department of Earth and Environmental Engineering, Columbia University, NY, 10027, USA.
E-mail: yc2965@columbia.edu

Qi Li
Department of Civil and Environmental Engineering, Cornell University, NY, 14853, USA.

Andrey Grachev
NOAA Earth System Research Laboratory / Cooperative Institute for Research in Environmental Sciences, University of Colorado, 325 Broadway, R/PSD3, Boulder, CO 80305-3337, USA

Stefania Argentini
Institute of Atmospheric Sciences and Climate, CNR, Rome, Italy.

Harindra Joe Fernando
Department of Civil and Environmental Engineering and Earth Sciences, University of Notre Dame, IN, 46556, USA.

range of the stable atmospheric boundary layer. This finding contributes to improved estimation of turbulent fluxes in both modelling and observation.

**1 Introduction**

Turbulence cospectra of atmospheric surface fluxes are typically assumed to follow a -7/3 power-law scaling in the universal equilibrium range according to a theoretical derivation based on dimensional analysis (Lumley 1964; Lumley 1967). The -7/3 power-law scaling has been observed by atmospheric measurements from the Kansas experiment (Kaimal et al. 1972; Wyngaard and Coté 1972). In eddy-covariance (EC) measurements of turbulent fluxes in the atmospheric surface layer, an assumed cospectral shape is required for spectral corrections of momentum, heat, water vapour and $CO_2$ fluxes (Moore 1986; Leuning and Moncrieff 1990; Horst 1997; Moncrieff et al. 1997; Aubinet et al. 1999; Massman 2000). Recently, Mamadou et al. (2016) showed that the calculated long-term $CO_2$ fluxes from EC observation are particularly sensitive to the assumed cospectra shape, and a change of shape can even reverse a net carbon sink to a source. In numerical calculation of atmospheric surface fluxes, the stability correction functions (Monin and Obukhov 1954; Kaimal et al. 1972; Stull 1988) are usually applied, which require calibration from EC observations, thus invoking an assumed cospectra shape.

Monin and Yaglom (1975) earlier pointed out that Lumley's derivation of a -7/3 scaling (Lumley 1964; Lumley 1967) was not sufficiently rigorous and accurate. Besides, a -2 scaling was then reported in a wind tunnel experiment (Mydlarski and Warhaft 1998). Bos et al. (2004) then suggested that velocity-scalar cospectra can in fact have a slope between -7/3 and -5/3 from spectral flux analysis. Here we report a -2 power-law scaling for turbulence cospectra at high wavenumbers in the stably stratified atmospheric boundary layer (ABL), which is supported by various high-quality field measurements.

## 2 Derivation of cospectrum scaling

The energy transfer rate due to turbulent heat flux in stratified turbulence is defined as (Richardson 1920)

$$\frac{g}{\theta_0}\langle w'\theta'\rangle, \qquad (1)$$

where $g$ is the acceleration of gravity, $\theta_0$ is the mean potential temperature in a horizontal plane, $w'$ is the vertical velocity fluctuation, $\theta'$ is the fluctuation of potential temperature and $\langle\ \rangle$ is averaging in a horizontal plane. In the ABL (Kaimal and Finnigan 1994), cospectrum is the real part of Fourier transform of cross-covariance. Here we have (Kaimal and Finnigan 1994)

$$\langle w'\theta'\rangle = \int_{-\infty}^{\infty} Co_{w\theta}(k)dk, \qquad (2)$$

where $Co_{w\theta}$ is the cospectrum of $\langle w'\theta'\rangle$, $k$ the wavenumber, $w$ the vertical velocity and $\theta$ the potential temperature. Assuming the cospectrum of heat flux is only related to potential temperature gradient $\frac{\partial \theta_0}{\partial z}$, turbulence dissipation rate $\epsilon$ and wavenumber $k$ for isotropic turbulence, Lumley (1964) obtained the following form for the cospectrum using a dimensional analysis

$$Co_{w\theta} = -c_1 \epsilon^{1/3} N^2 k^{-7/3}, \qquad (3)$$

where $c_1$ is a constant, $N^2 = \frac{g}{\theta_0}\frac{\partial \theta_0}{\partial z}$ and $N$ is Brunt–Väisälä frequency.

Similarly, Lumley (1967) suggested the cospectrum of momentum flux in the form:

$$Co_{wu} = -c_2 \epsilon^{1/3} \frac{\partial U}{\partial z} k^{-7/3}, \qquad (4)$$

where $c_2$ is a constant, $U$ the mean streamwise velocity in a horizontal plane and $u$ the streamwise velocity. The above derivation are based on dimensional analysis so there could be other slopes for stratified and sheared flows, for example,

$$Co_{w\theta} = -c_3 \epsilon^{1/2} N^{3/2} k^{-2}, \qquad (5)$$

and

$$Co_{wu} = -c_4 \epsilon^{1/2} \left(\frac{\partial U}{\partial z}\right)^{3/2} k^{-2}, \qquad (6)$$

where $c_3$ and $c_4$ are constants. So a -2 scaling could be valid within dimensional analysis. Other scalar fluxes such as water vapor, $CO_2$ are assumed to have the same scaling as heat flux in the equilibrium range.

We emphasize that the above scaling analysis is only strictly applicable for isotropic turbulence (Kolmogorov 1941). It is generally believed that the Dougherty-Ozmidov scale (Dougherty 1961; Ozmidov 1965; Gargett et al. 1984) $L_O = 2\pi \left(\frac{\epsilon}{N^3}\right)^{\frac{1}{2}}$ named in Grachev et al. (2015) is the largest scale of (isotropic turbulence) in stably stratified fluid. The corresponding wavenumber to $L_O$ is $k_O = \frac{2\pi}{L_O}$. Due to wall effects (Townsend 1976; Katul et al. 2014) in the ABL, the wavenumber $k_a$ corresponding to wall-attached eddies, which satisfies $k_a z = 1$ will also constrain the existence of isotropic turbulence, where $z$ is the height above ground. Therefore, similarly to the turbulent kinetic energy (TKE) spectra (Cheng et al. 2018), we expect an isotopic turbulence power-law scaling for cospectra at wavenumber $k > \max(k_O, k_a)$. As suggested by Waite (2011), the equilibrium range also includes the buoyancy scale $L_b = 2\pi \frac{U_{RMS}}{N}$ (Billant and Chomaz 2001), where $U_{RMS}$ is root mean square of the horizontal velocity. We will show the impact of that wavenumber $k_b = \frac{2\pi}{L_b}$ on some of the cospectra.

**3 Experiment setup and results**

3.1 Observations of the stable atmospheric boundary layer

An eddy-covariance (EC) system over Lake Geneva was set up to measure high-frequency (20 Hz) velocity, temperature, water vapour and $CO_2$ concentration at 4 different heights (1.66 m, 2.31 m, 2.96 m and 3.61 m above water level) during August-October 2006 (Bou-Zeid et al. 2008). Four sonic anemometers (Campbell Scientific CSAT3) and open-path gas analyzers (LICOR LI-7500) were used in the experiment. The resolution of wind velocity was 0.001 m s$^{-1}$ from instrument and that of temperature was 0.002 ℃. Nine representative 15-minute periods of EC data were selected to calculate turbulence cospectra of heat, momentum, water vapour and $CO_2$ fluxes. The reader is referred to (Bou-Zeid et al. 2008; Vercauteren et al. 2008; Li and Bou-Zeid 2011; Li et al. 2018) for detailed description of the experiment setup and data.

An EC system at Dome C, Antarctica was set up to measure high-frequency (10 Hz) velocity and temperature using an ultrasonic anemometer (Metek USA-1) at 3.5 m above ground (Vignon et al. 2017a; Vignon et al. 2017b). Balloon sounding measurements provided temperature gradient (Petenko et al. in press). Two 30-minute periods around 8 PM on 9 and 10 January 2015 were selected respectively. The accuracy of wind speed was 0.05 m s$^{-1}$ and that of temperature was 0.01 ℃. The reader is referred to (Vignon et al. 2017a) for details of experiment setup.

An EC system over an Arctic pack ice during the Surface Heat Budget of the Arctic Ocean experiment (SHEBA) was set up to measure high-frequency (10 Hz) velocity and temperature using ATI (Applied Technologies, Inc) three-axis sonic anemometer at 5 heights (2.2 m, 3.2 m, 5.1 m, 8.9 m, and 18.2 m or 14 m during most of winter) from October 1997 through September 1998 (Andreas et al. 2006; Grachev et al. 2013). The resolution of wind velocity was 0.01 m s$^{-1}$ from instrument and that of temperature was 0.01 ℃. Nine representative 60-minute periods of EC data were selected for cospectra of heat and momentum fluxes. The mean of turbulence cospectra of a few 13.65-minute periods in one hour were analyzed (Persson et al. 2002). The periods are denoted by the day of a year and the hour of a day, for example, "312h23" denotes the 23rd hour on the 312th day of a year. The reader is referred to detailed introduction (Grachev et al. 2005; Andreas et al. 2006; Andreas et al. 2010a; Andreas et al. 2010b; Grachev et al. 2013) for experiment setup and data.

A sonic and hot-film anemometer dyad (Kit et al. 2017) was installed at the Granite Mountain Atmospheric Sciences Testbed (GMAST) of the US army Dugway Proving Ground (DPG), Utah, in field measurements of the Mountain Terrain Atmospheric Modeling and Observations (MATERHORN) program during September-October 2012 (Fernando et al. 2015) to capture fine-scale turbulence in the ABL. Wind velocity was measured at a height of 2 m on a 32 m high tower with a temporal frequency of 2000 Hz. The resolution of the composite probe was ~0.7 mm, and the measurement resolution of hot-film X-wire probes was ~1mm. Nine representative 30-minute periods were selected for the cospectra of momentum fluxes on 9 October 2012. The reader is referred to detailed introduction of instrument setup and measurement method (Fernando et al. 2015; Kit and Liberzon 2016; Kit et al. 2017; Sukoriansky et al. 2018).

3.2 Turbulence cospectra

The Obukhov length $L = -\frac{u_*^3}{\frac{\kappa g}{T_0}\langle w'T'\rangle}$ (Obukhov 1946) was calculated to characterize the stability of the ABL, where $u_*$ is the friction velocity, $\kappa$ the von Kármán constant, $T_0$ the mean air temperature and $T'$ the fluctuation. We also used the horizontal Froude number $Fr = \frac{U_{RMS}}{NL_h}$ as a measure of the strength of stratification, where $U_{RMS}$ is root mean square of the horizontal streamwise velocity and $L_h = \frac{U_{RMS}^3}{\epsilon}$ (Lindborg 2006) is a horizontal length scale calculated from Taylor's frozen turbulence hypothesis (Taylor 1938). The flux Richardson number $R_f = \frac{g}{T_0}\frac{\langle w'T'\rangle}{\langle w'u'\rangle\frac{\partial U}{\partial z}}$ was used to characterize the relative importance of buoyancy flux and production of TKE. Rather than directly measuring cospectra in wavenumber space, we converted frequency cospectra into wavenumber cospectra, by invoking Taylor's frozen turbulence hypothesis. Wavelet transform (Torrence and Compo 1998) was used to calculate turbulence cospectra (software was provided by C. Torrence and G. Compo, and is available at: http://paos.colorado.edu/research/wavelets/).

Both frequency and cospectra were normalized in a similar way to Kaimal et al. (1972). Nine representative 15-minute periods in the stable ABL were selected in EC measurements over Lake Geneva (Fig. 1). A -2 slope for cospectra of heat flux at wavenumber $k > \max(k_O, k_a)$ is evidenced in each period (Fig. 1). A few cospectra jumped at the highest wavenumbers are seen in some periods due to limitation of instrument sampling, which can also be better observed in the compensated cospectra (Fig. 2). At low wavenumbers, the cospectra slope is shallower than -2, and even approaches zero in some periods. Because internal gravity waves (Lumley 1964; Caughey and Readings 1975; Smedman 1988) have stronger impacts on larger eddies, turbulence deviates more from isotropic condition at lower wavenumbers (Lienhard and Van Atta 1990). It can also be seen that cospectral scaling deviates more from the -2 scaling at lower wavenumbers. The cospectrum is negative (heat flux is downward to the ground) in most part of wavenumber space but is positive (flux is upward i.e. counter-gradient) around the lowest wavenumbers. Besides, in most periods the upward flux occur at scales larger than the buoyancy scale. The upward flux is counter-gradient as has been observed in direct numerical simulations (DNS) of stably stratified sheared turbulence (Holt et al. 1992) and experimental studies (Schumann 1987; Komori and Nagata 1996), which can be interpreted as a mechanism converting potential energy to kinetic energy (Schumann 1987) at high Richardson numbers. Komori and Nagata (1996) further suggested that counter-gradient heat flux only appears at low wavenumbers in the air even if total heat flux is upward in strong stratification conditions (Lienhard and Van Atta 1990; Yoon and Warhaft 1990). This is confirmed by our experiential data as this counter-gradient flux is mainly due to low wavenumber eddies.

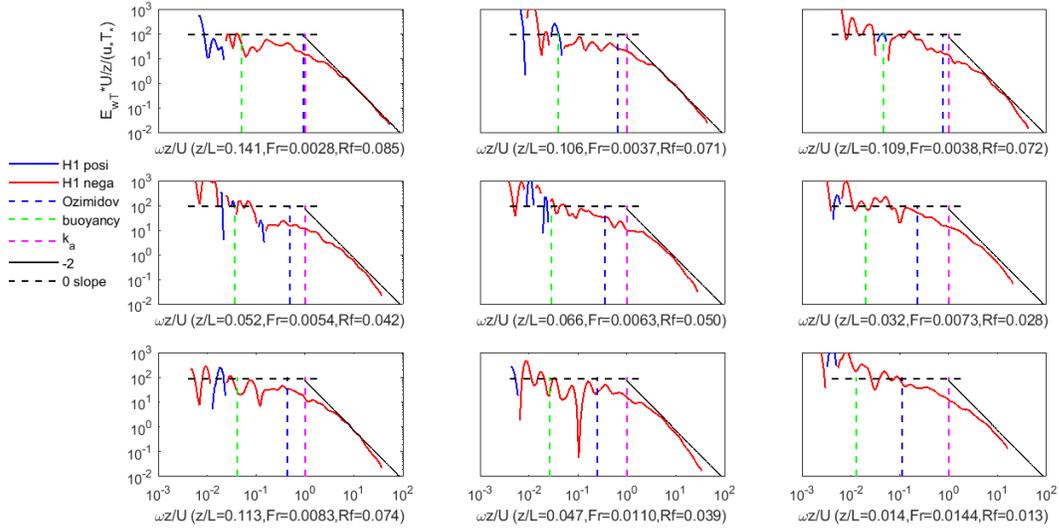

**Fig.1** Normalized temporal wavelet cospectra of heat flux in 9 representative 15-minute periods of Lake Geneva eddy-covariance data. $E_{wT}$ is the wavelet cospectrum of the vertical velocity fluctuations $w$ and temperature fluctuation $T$ in time, $U$ the mean streamwise wind velocity, $z$ the measurement height above the lake, $u_*$ the friction velocity, $T_*$ the scaling temperature, $\omega = 2\pi f$ is angular frequency, $f$ is sampling frequency in Hz, $L$ the Obukhov length, $Fr$ the horizontal Froude number and $R_f$ the flux Richardson number. "H1 posi" denotes positive flux at height 1.66m above the lake and "H1 nega" denotes negative flux at the same height. "Ozmidov", "buoyancy", "$k_a$" denotes the Dougherty-Ozmidov scale (named in Grachev et al. (2015)), the buoyancy scale and "$k_a$" respectively.

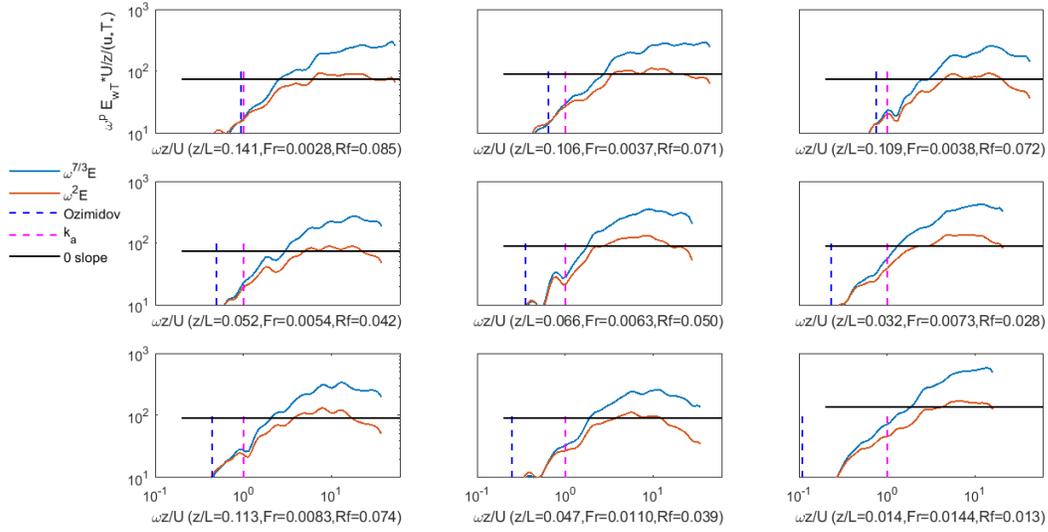

**Fig. 2** Normalized temporal cospectra of heat flux (denoted by $E$) multiplied by $\omega^{7/3}$ (blue lines) or $\omega^2$ (red lines) in 9 representative 15-minute periods of Lake Geneva EC data. $p$ is an exponent equal to 7/3 or 2 and other variables have the same meaning as those in Fig. 1.

To further check whether a -2 or -7/3 is the asymptotic slope for cospectra at high wavenumbers, the normalized cospectra of heat flux were multiplied by $\omega^2$ and $\omega^{7/3}$ respectively (Fig. 2), where $\omega = 2\pi f$ is angular frequency and $f$ is sampling frequency in Hz. A positive slope can be observed at wavenumber $k > \max(k_O, k_a)$ for $\omega^{7/3} E_{wT}$ in most periods, where $E_{wT}$ denotes the heat flux cospectrum. However, a plateau is more obvious and occupies more wavenumber bandwidth when using $\omega^2 E_{wT}$, further confirming that the high-frequency scaling is in -2 power. Note that the plateau does not appear immediately at $k > \max(k_O, k_a)$. Therefore, we conclude based on this dataset that the -2 scaling is more appropriate than -7/3 at high wavenumbers for the heat cospectrum.

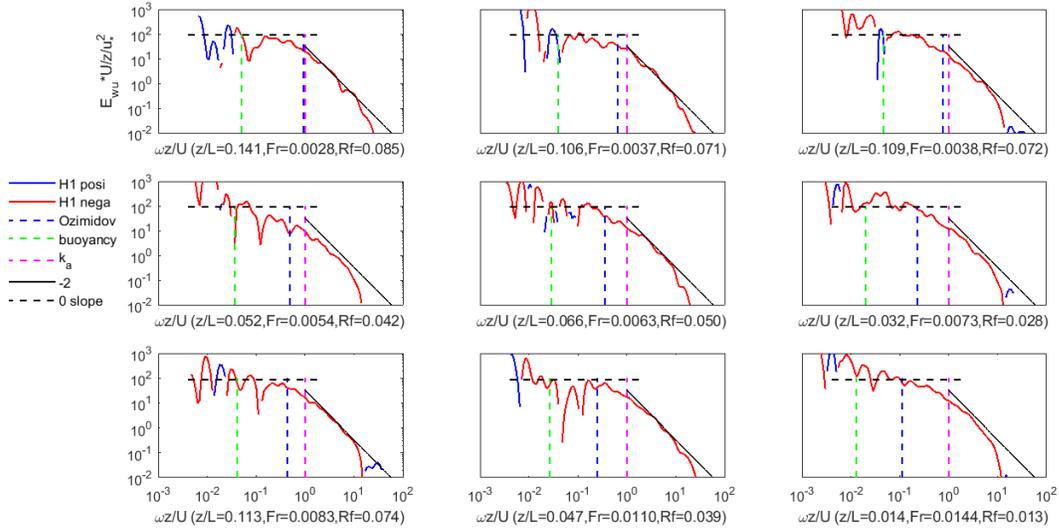

**Fig. 3** Normalized temporal cospectra of momentum flux in 9 representative 15-minute periods of Lake Geneva EC data. $E_{wu}$ is wavelet cospectrum of vertical velocity fluctuation $w$ and streamwise velocity fluctuation $u$ in time. Other variables have the same meaning as those in Fig. 1.

Similarly to heat flux, a -2 scaling is observed at wavenumber $k > \max(k_O, k_a)$ in the cospectra of momentum (Fig. 3), water vapour (Fig. 5) and $CO_2$ (Fig. 7) flux for the same 9 periods. Compared to the cospectra of heat flux, there is a more obvious cospectra jump at the highest wavenumber in momentum, water vapour and $CO_2$ flux due to high-frequency attenuation (Moore 1986; Moncrieff et al. 1997). Counter-gradient transfer also exists especially around lowest wavenumbers in the cospectra of those fluxes. At wavenumber $k > \max(k_O, k_a)$, the compensated $\omega^{7/3} E_{wu}$ ($E_{wu}$ denotes cospectrum of momentum flux) shows a positive slope in most periods but $\omega^2 E_{wu}$ exhibits a larger plateau in wavenumber space (Fig. 4). That is also the case for water vapour: $\omega^{7/3} E_{wq}$ does not exhibit a plateau compared to $\omega^2 E_{wq}$ (Fig. 6) and for $CO_2$ as $\omega^{7/3} E_{wC}$ exhibits a slight slope compared to $\omega^2 E_{wC}$ (Fig. 8). $E_{wq}$ denotes the cospectra of water vapor flux and $E_{wC}$ denotes the cospectra of $CO_2$ flux. Compared to other fluxes, $CO_2$ and water vapour fluxes have more noise in its cospectra due to instrument noise and sampling of water vapour (Webb et al. 1980; Sahlée et al. 2008; Oikawa et al. 2017).

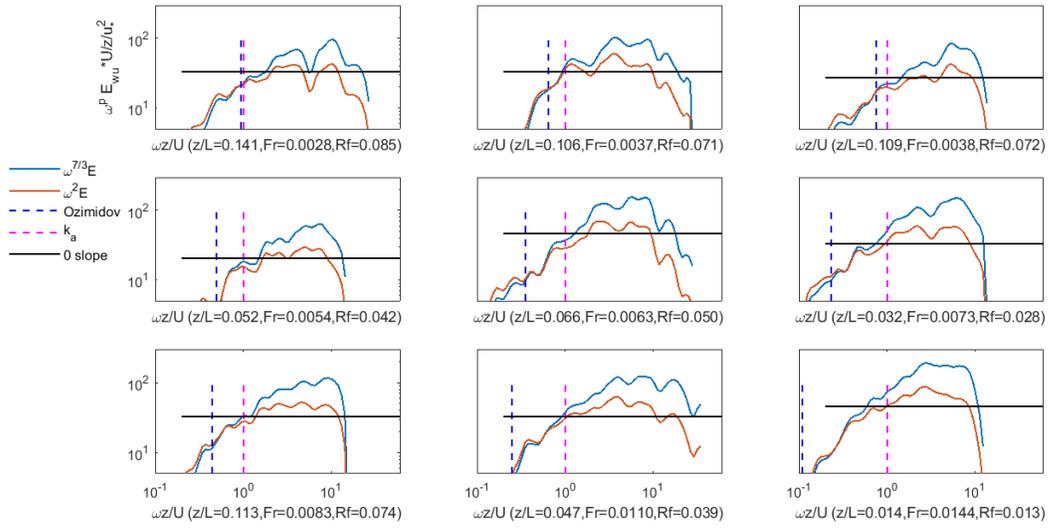

**Fig. 4** Normalized temporal cospectra of momentum flux (denoted by $E$) multiplied by $\omega^{7/3}$ or $\omega^2$ in 9 representative 15-minute periods of Lake Geneva EC data. $E_{wu}$ is wavelet cospectrum of vertical velocity fluctuation $w$ and streamwise velocity fluctuation $u$ in time, $p$ is an exponent equal to 7/3 or 2 and other variables have the same meaning as those in Fig. 1.

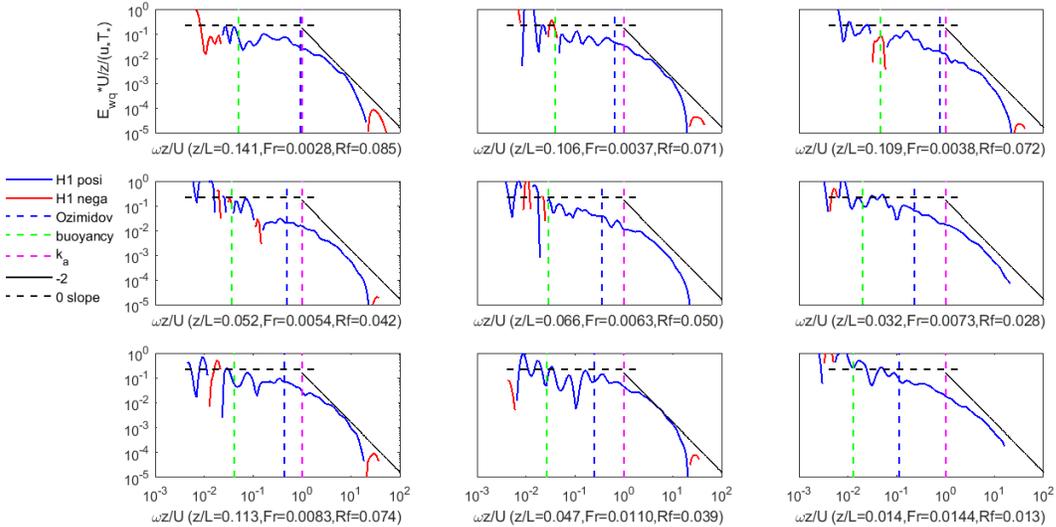

**Fig. 5** Normalized temporal cospectra of water vapor flux in 9 representative 15-minute periods of Lake Geneva EC data. $E_{wq}$ is wavelet cospectrum of vertical velocity fluctuation $w$ and water vapor fluctuation $q$ in time. Other variables have the same meaning as those in Fig. 1.

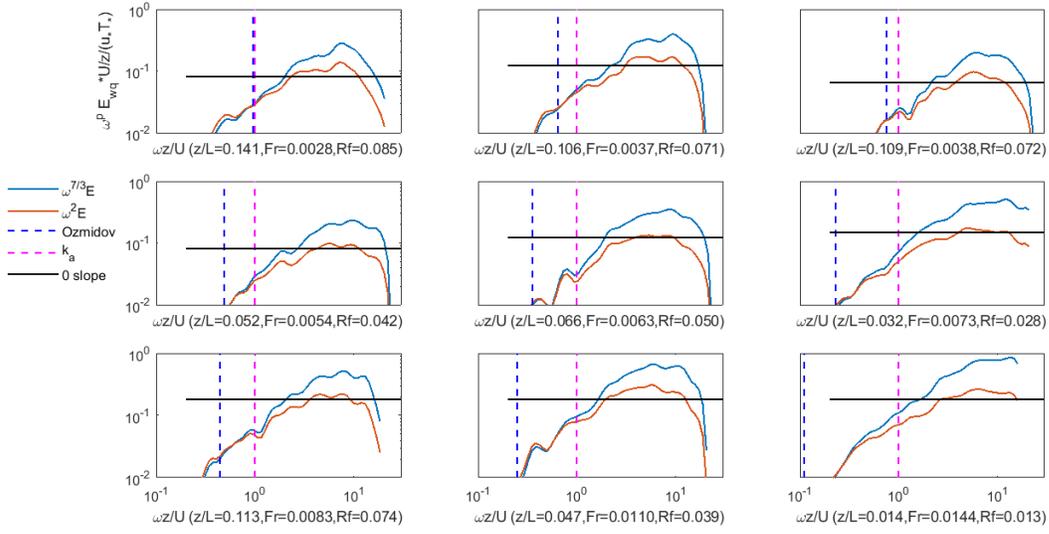

**Fig. 6** Normalized temporal cospectra of water vapour flux (denoted by $E$) multiplied by $\omega^{7/3}$ or $\omega^2$ in 9 representative 15-minute periods of Lake Geneva EC data. $E_{wq}$ is wavelet cospectrum of vertical velocity fluctuation $w$ and water vapour fluctuation $q$ in time, $p$ is an exponent equal to 7/3 or 2 and other variables have the same meaning as those in Fig. 1.

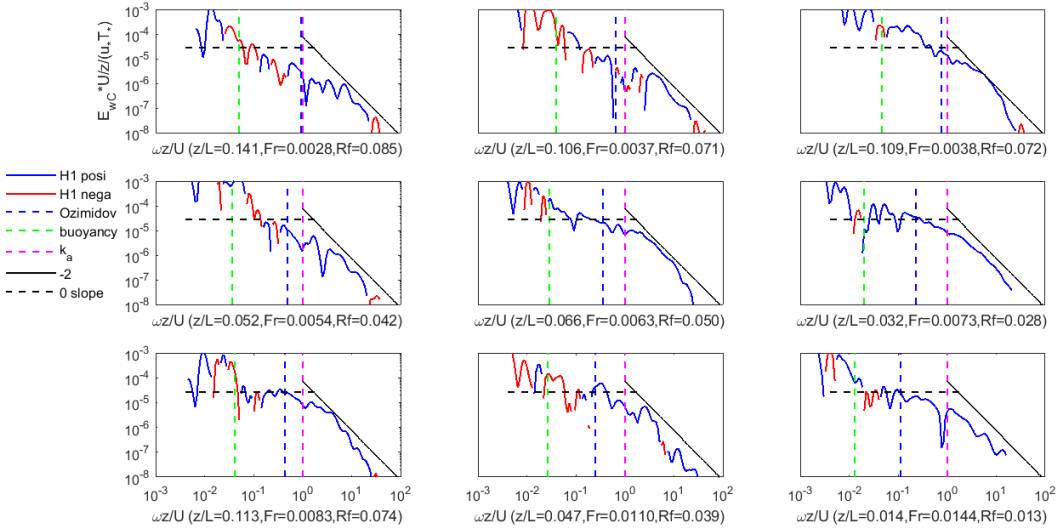

**Fig. 7** Normalized temporal cospectra of $CO_2$ flux in 9 representative 15-minute periods of Lake Geneva EC data. $E_{wc}$ is wavelet cospectrum of vertical velocity fluctuation $w$ and $CO_2$ fluctuation $C$ in time. Other variables have the same meaning as those in Fig. 1.

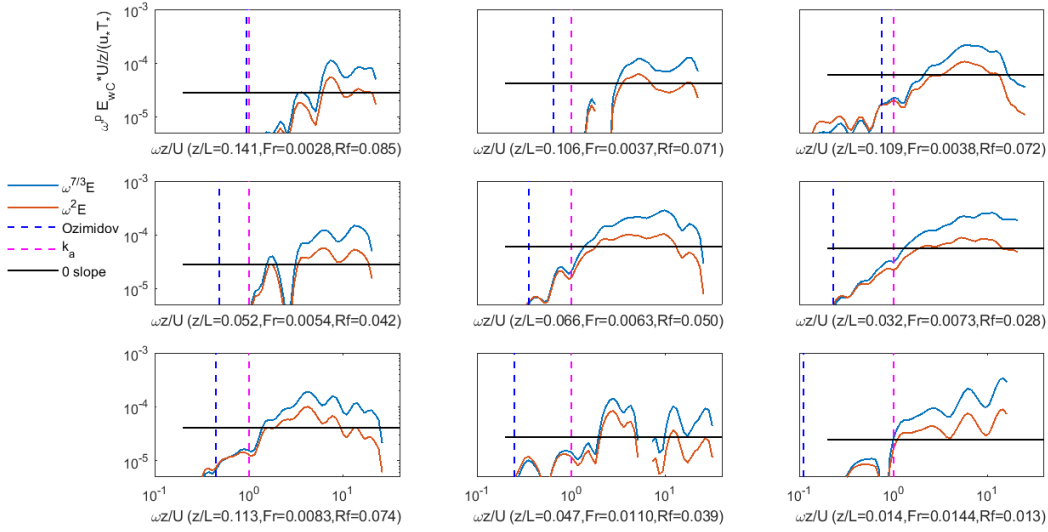

**Fig. 8** Normalized temporal cospectra of $CO_2$ flux (denoted by $E$) multiplied by $\omega^{7/3}$ or $\omega^2$ in 9 representative 15-minute periods of Lake Geneva EC data. $E_{wC}$ is wavelet cospectrum of vertical velocity fluctuation $w$ and $CO_2$ fluctuation $C$ in time, $p$ is an exponent equal to 7/3 or 2 and other variables have the same meaning as those in Fig. 1.

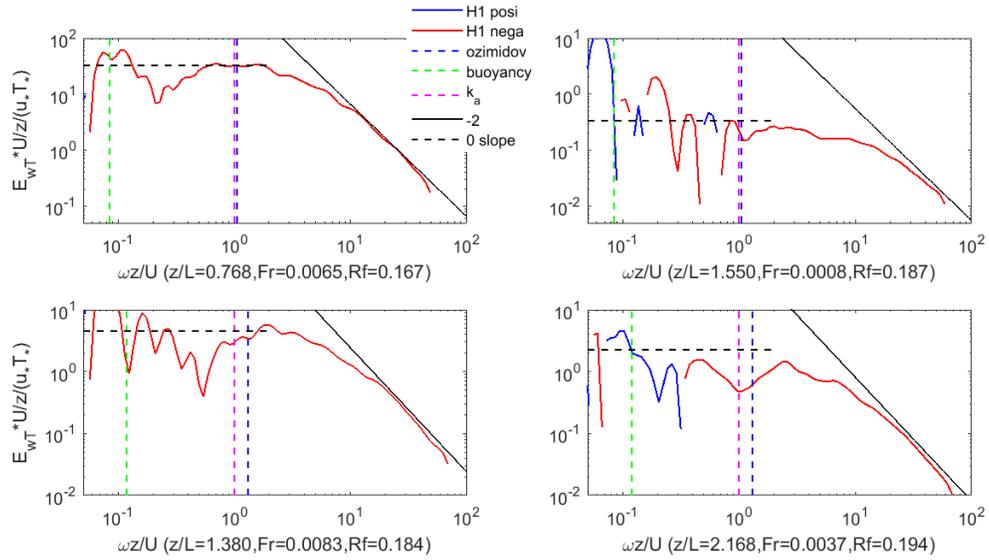

**Fig. 9** Normalized temporal cospectra of heat flux in 4 representative 30-minute periods of Dome C EC data. Variables have the same meaning as those in Fig. 1.

Four representative 30-minute periods in the stable ABL were selected for EC at Dome C in Antarctica (Figs. 9 & 11). A -2 slope for cospectra of heat and momentum flux is clearly seen at wavenumber $k > \max(k_O, k_a)$ in each period covering about one decade. Similarly to the Lake data, $\omega^{7/3} E_{wT}$ ($\omega^{7/3} E_{wu}$) shows a positive slope but $\omega^2 E_{wT}$ ($\omega^{7/3} E_{wu}$) has a larger plateau, as shown in Fig. 10 (Fig. 12) at high wavenumbers.

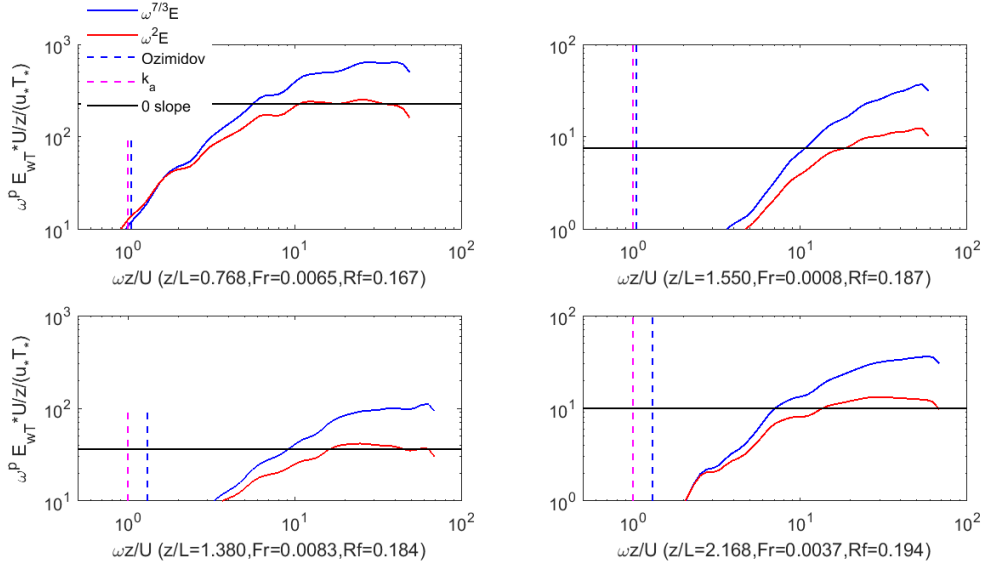

**Fig. 10** Normalized temporal cospectra of heat flux (denoted by $E$) multiplied by $\omega^{7/3}$ or $\omega^2$ in 4 representative 30-minute periods of Dome C EC data. $p$ is an exponent equal to 7/3 or 2 and other variables have the same meaning as those in Fig. 1.

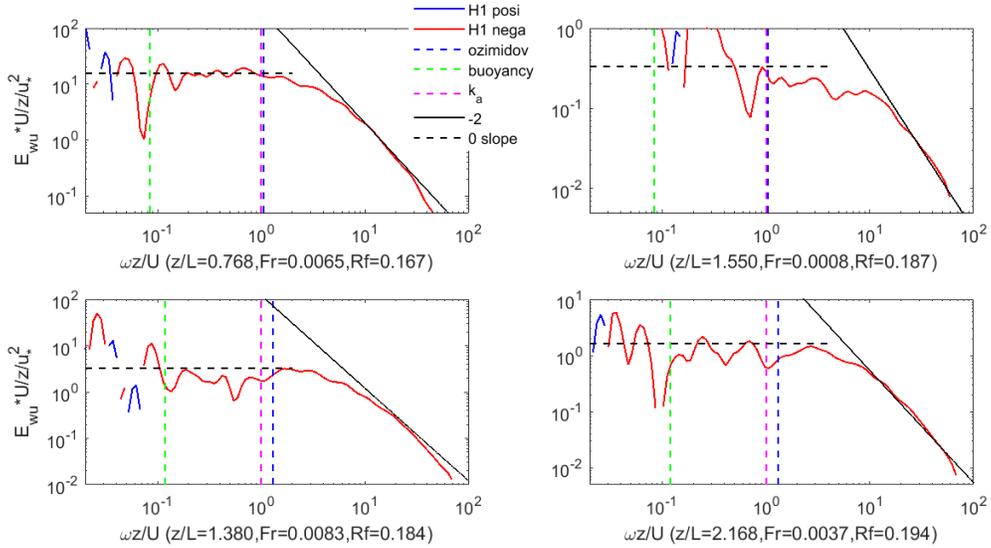

**Fig. 11** Normalized temporal cospectra of momentum flux in 4 representative 30-minute periods of Dome C EC data. $E_{wu}$ is wavelet cospectrum of vertical velocity fluctuation $w$ and streamwise velocity fluctuation $u$ in time. Other variables have the same meaning as those in Fig. 1.

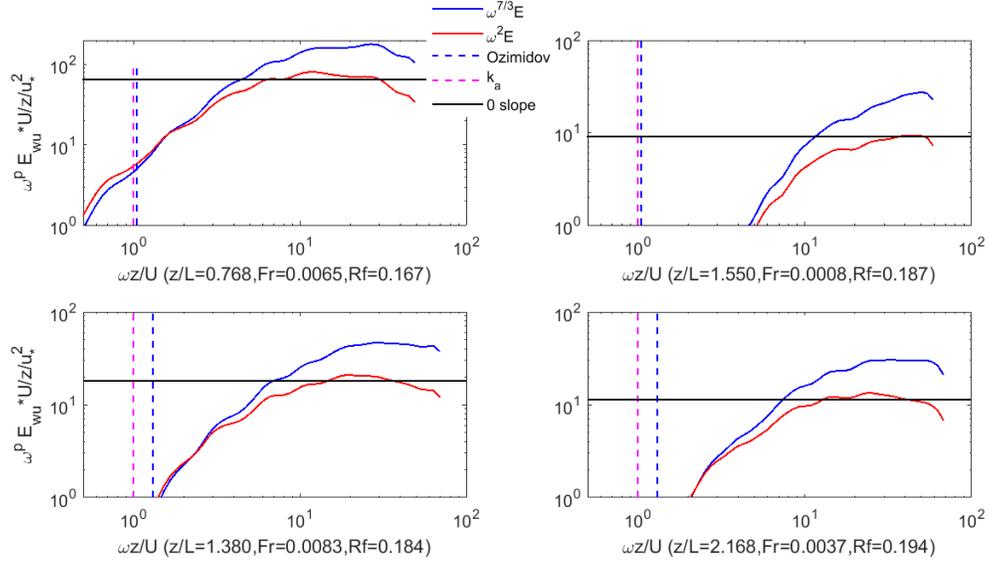

**Fig. 12** Normalized temporal cospectra of momentum flux (denoted by $E$) multiplied by $\omega^{7/3}$ or $\omega^2$ in 4 representative 30-minute periods of Dome C EC data. $E_{wu}$ is wavelet cospectrum of vertical velocity fluctuation $w$ and streamwise velocity fluctuation $u$ in time. $p$ is an exponent equal to $7/3$ or $2$. Other variables have the same meaning as those in Fig. 1.

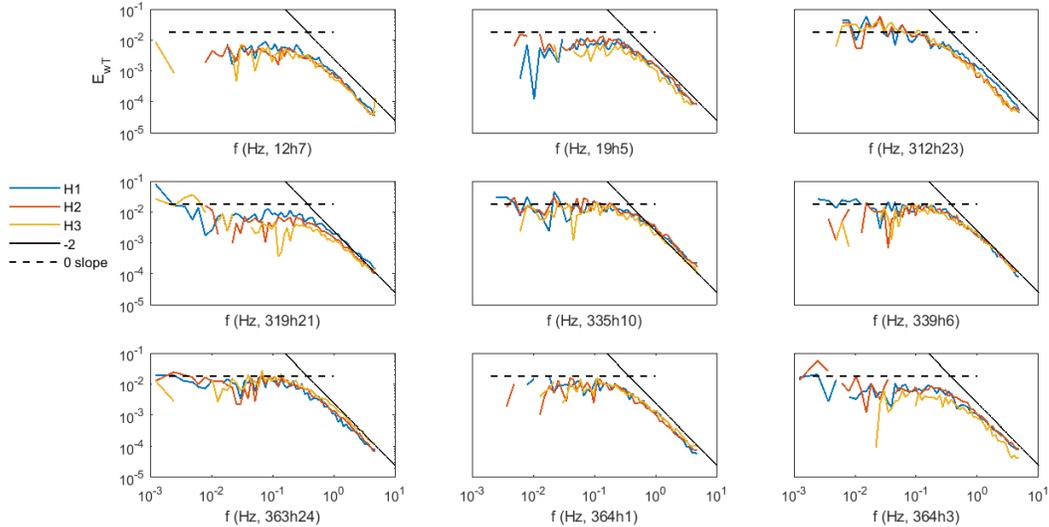

**Fig. 13** Temporal cospectra of heat flux (denoted by $E_{wT}$) in 9 representative 13.65-minute periods of SHEBA EC data. $E_{wT}$ is wavelet cospectrum of vertical velocity fluctuation $w$ and temperature fluctuation $T$ in time, $f$ is sampling frequency in Hz, "12h7" denotes 7th hour of 12th day in a year and other periods follow the same rule. "H1", "H2" and "H3" denotes measurement heights at 2.2 m, 3.2 m and 5.1 m above ice respectively.

Nine representative 1-hour averaged periods in the stable ABL were selected for EC at SHEBA campaign over Arctic Sea to evaluate the heat flux cospectra (Fig. 13). The cospectra were calculated from overlapping 13.65-minute blocks (corresponding $2^{13}$ data points) and then averaged over 1-hour period (Persson et al. 2002). Cospectra are plotted against frequency in each periods (Fig. 13). A -2 slope for the cospectra of heat flux at high wavenumbers is again evident, covering at least a decade, with minimal high-frequency noise. $f^{7/3}E_{wT}$ shows a positive slope in most periods but $f^2 E_{wT}$ has a larger plateau as shown in Fig. 14 at high wavenumbers.

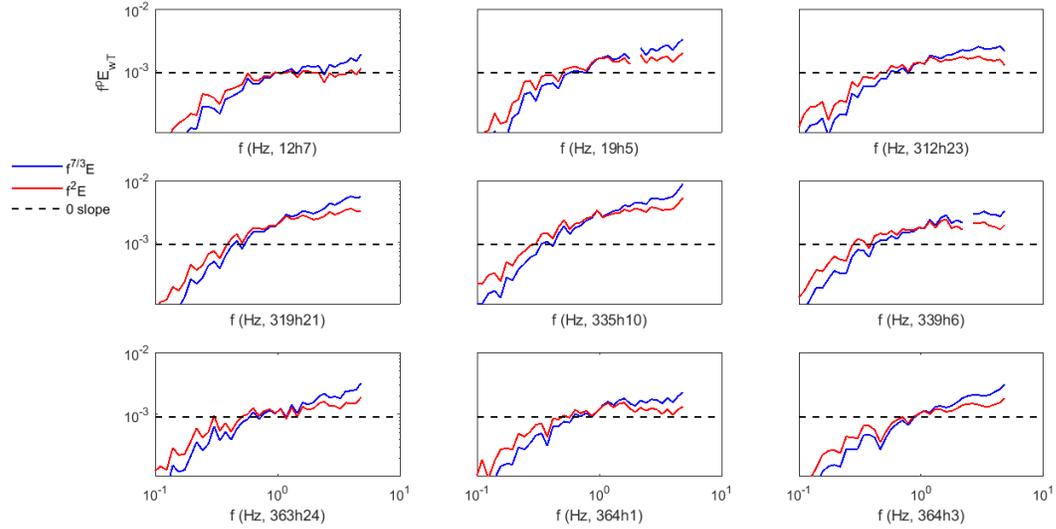

**Fig. 14** Temporal cospectra of heat flux (denoted by $E$ or $E_{wT}$) multiplied by $f^{7/3}$ or $f^2$ in 9 representative averaged 13.65-minute periods of SHEBA EC data. $p$ is an exponent equal to 7/3 or 2. Other variables have the same meaning as those in Fig. 13.

Nine representative 30-minute periods in the stable ABL were then selected for the MATERHORN campaign (Fig. 15), which provides unique very high-frequency data up to 400Hz, thus allowing one to further test the validity of the power law slope at very high frequencies. The cospectra of momentum flux calculated based on the fast Fourier transform (Frigo and Johnson 1998) were multiplied by $f^{7/3}$ or $f^2$. Although there are a spikes in the compensated cospectra due to measurement noise, $f^{7/3}E_{wu}$ shows a positive slope from 40 Hz to 400 Hz in non-spike part in most periods while $f^2 E_{wu}$ is mostly flat in the same frequency domain. This further confirms that a -2 slope appears to be universal for the cospectra of the stable boundary layer. Therefore, the -2 scaling matches the ABL observation better than the -7/3 scaling in various stratified atmospheric conditions.

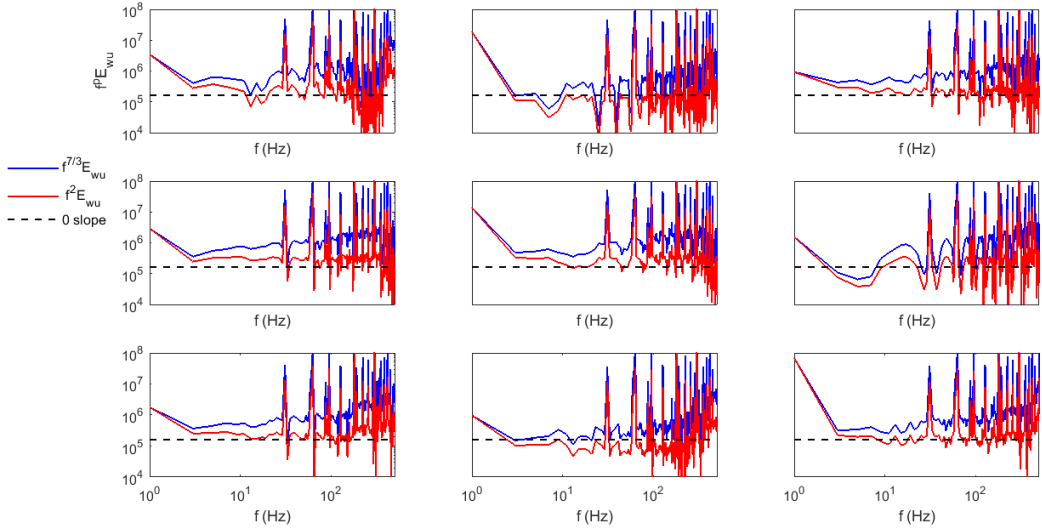

**Fig. 15** Temporal Fourier cospectra of momentum flux multiplied by $f^{7/3}$ or $f^2$ in 9 representative 30-minute periods of MATERHORN hot film data. $E_{wu}$ is wavelet cospectrum of vertical velocity fluctuation $w$ and streamwise velocity fluctuation $u$ in time, $f$ is sampling frequency in Hz and $p$ is an exponent equal to 7/3 or 2.

3.3 Discussion

In a wind tunnel experiment, an asymptotic -2 scaling was observed at high wavenumbers of the compensated cospectra of heat flux (Mydlarski and Warhaft 1998) at $R_\lambda = 582$ ($R_\lambda$, Taylor-microscale-based Reynolds number) in stably stratified turbulence. Mydlarski (2003) also found a -2 scaling in heat flux cospectra through analysis of both the cospectra and the heat flux structure function (Mydlarski 2003) at $R_\lambda = 407$ when a temperature gradient was imposed in the transverse direction, although the paper suggested that the slope might increase toward -7/3. Sakai et al. (2008) showed a -2 scaling for radial velocity-concentration cospectra in a turbulent jet. Among numerical studies, O'Gorman and Pullin (2005) found a scaling close to -2 in velocity-scalar cospectra in DNS of homogeneous and isotropic velocity field mixed with a mean scalar gradient. Watanabe and Gotoh (2007) observed a -2 scaling region to the right side of a -7/3 scaling region in the cospectrum of scalar flux with high resolution DNS at $R_\lambda = 585$. In fact, figure 2 in their paper clearly shows that the -2 scaling has a larger plateau compared to the -7/3 scaling in the compensated cospectra so -2 appears to be a better approximation to the cospectrum slope. Bos et al. (2004) also found a clear -2 scaling in velocity-scalar cospectra in large eddy simulations (LES) with a mean scalar gradient. It is worth noting that some studies (Kaimal et al. 1972; Saddoughi and Veeravalli 1994; Bos 2014) reported a -7/3 scaling for cospectra but did not compare data with -2 scaling, which is still close to -7/3 and could thus be mixed. Therefore, both laboratory and numerical experiments primarily support a -2 scaling rather than a -7/3 scaling in the turbulence cospectra of heat, momentum and scalar flux at moderate Reynolds number ($R_\lambda \leq 10^3$), which are consistent with our field observations in the stable ABL ($R_\lambda \sim 10^4$) at low Froude number.

In terms of theoretical analyses, O'Gorman and Pullin (2003) proposed that both a -5/3 scaling leading term and a next order -7/3 scaling term contribute to the cospectra of axial velocity and scalar based on a stretched-spiral vortex model. Bos et al. (2005) showed from an eddy-damped quasi-normal Markovian (EDQNM) (Orszag 1970) closure that a -7/3 scaling for scalar cospectra could only be observed at very high Reynolds number ($R_\lambda = 10^7$) while a shallower cospectra could be observed at relatively lower Reynolds number. In other words, these theoretical models suggest that the -2 scaling could be observed at moderate Reynolds number ($R_\lambda \sim 10^4$) such as in the ABL (Bradley et al. 1981; Gulitski et al. 2007). Considering our ABL observations, previous numerical simulations and theoretical modeling, it is reasonable to conclude that -2 rather than -7/3 is a better approximation of cospectral scaling in the stratified ABL at $R_\lambda \sim 10^4$. As for turbulence spectra, the asymptotic -5/3 scaling appears at $R_\lambda = 600$ (Saddoughi and Veeravalli 1994) in velocity spectra and appears at $R_\lambda = 731$ in temperature spectra (Mydlarski and Warhaft 1998). It is reasonable to assume that the asymptotic cospectral scaling also appears at $R_\lambda$ on the order of 600 in a similar way to spectra. Therefore, the -2 scaling observed at $R_\lambda \sim 10^4$ might be the asymptotic cospectral scaling at infinite Reynolds number.

Equation (33) in Kaimal et al. (1972) suggested a "-2.1" power-law scaling for the cospectra of heat and momentum fluxes at high wavenumbers, while claiming -7/3 slope as the asymptotic cospectral scaling. The "-2.1" slope was then adopted in a spectral correction method (Moore 1986). Horst (1997) assumed a -2 scaling for scalar flux as it approximates observations, and could be analytically computed. Massman (2000) and Massman and Lee (2002) also suggested -2 scaling for spectral correction of EC measurements. Therefore, the -2 scaling has been applied in some earlier spectral correction methods of EC measurements, yet without strong justification. Here we provide multiple lines of evidence emphasizing that the cospectra should obey a -2 scaling in the stable ABL.

## 4 Conclusion

We observed a -2 power-law scaling of turbulence cospectra across various field measurements in the stably stratified ABL. The observations are consistent with moderate Reynolds number ($R_\lambda \leq 10^4$) results of laboratory experiments (Mydlarski and Warhaft 1998; Mydlarski 2003; Sakai et al. 2008), DNS (O'Gorman and Pullin 2005; Watanabe and Gotoh 2007) and LES (Bos et al. 2004) studies, which compared the -2 scaling with the -7/3 scaling. Although it remains unknown what could be the asymptotic turbulence cospectra at infinite Reynolds number, we can infer from our observations that -2 rather than -7/3 is a better approximation of the cospectrum scaling in the stably stratified ABL and at moderate Reynolds number (with low Froude number). Therefore, a -2 power-law scaling should be applied in spectral corrections of EC measurements. The proposed power-law scaling should improve surface flux estimation especially when fluxes are sensitive to the assumed shape of turbulent cospectra such as those for $CO_2$ fluxes (Mamadou et al. 2016).

**Acknowledgements** PG would like to acknowledge funding from the National Science Foundation (NSF CAREER, EAR-1552304), and from the Department of Energy (DOE Early Career, DE-SC00142013). The lake data were collected by the Environmental Fluid Mechanics and Hydrology Laboratory of Professor M. Parlange at L'Ecole Polytechnique Fédérale de Lausanne. We would like to thank Prof. M. Parlange and Prof. Elie Bou-Zeid for sharing Lake EC data. Mountain Terrain Atmospheric Modeling and Observations (MATERHORN) program was funded by the Office of Naval Research (N00014-11-1-0709). Dome C data were acquired in the frame of the projects Mass lost in wind flux (MALOX) and Concordia multi-process atmospheric studies (COMPASS) sponsored by PNRA. A special thanks to P. Grigioni and all the staff of Antarctic Meteorological Observatory of Concordia for providing the radio sounding used in this study. And a special thanks to Dr. Igor Petenko of CNR ISAC for running the field experiment at Concordia station.